\documentclass{aa}
\usepackage[dvips]{graphicx}
\usepackage{psfig}

\newcommand{\lapp}{\mbox{\raisebox{-0.3em}{$\stackrel{\textstyle <}{\sim}$}}}
\newcommand{\gapp}{\mbox{\raisebox{-0.3em}{$\stackrel{\textstyle >}{\sim}$}}}

\begin{document}

\title{Polarization asymmetry in CSS sources: evidence of AGN fuel?}

\author{ D.J. Saikia \and Neeraj Gupta }

\titlerunning{Polarization asymmetry in CSSs}

\offprints{D.J. Saikia}

\institute{
National Centre for Radio Astrophysics, TIFR, Post Bag 3, Ganeshkhind, Pune 411 007, India
}

\date{Received 0000; Accepted 0000}

\abstract{
The compact steep spectrum and gigahertz peaked spectrum sources are widely 
believed to be young radio sources, with ages $\lapp$10$^6$ yr. If the activity in the
nucleus is fuelled by the supply of gas, one might find evidence of this gas by
studying the structural and polarization characteristics of CSS sources as these evolve
through this gas. We present polarization observations of a sample
of CSS sources, and combine our results with those available in the literature, to show
that CSS sources are more asymmetric in the polarization of the outer lobes compared
with the more extended ones.  We suggest that this could be possibly due
to interaction of the jets with infalling material, which fuels the radio source. 
We also investigate possible dependence of the polarization asymmetry of the lobes on 
redshift, since this might be affected by more interactions and mergers in the past. 
No such dependence is found for the CSS sources, suggesting that the environments on the
CSS scales are similar at different redshifts. However, the polarization asymmetry of the 
oppositely-directed lobes is larger at higher redshifts for the more extended sources,
possibly reflecting the higher incidence of interactions in the past. 
}

\maketitle

\section{Introduction }
There is a  consensus of opinion that the compact steep-spectrum (CSS) and 
gigahertz peaked spectrum (GPS) sources, defined to be $\leq$20 kpc in a Universe with 
q$_o$=0 and H$_o$=100 km s$^{-1}$ Mpc$^{-1}$, are young objects seen at an
early stage of their evolution.
Recent measurements of component advance speeds for a few very compact sources yield ages 
of about 10$^3$ yr (Taylor et al.\ 2000; Polatidis \& Conway 2003), while spectral studies 
of CSS sources suggest ages $\lapp$10$^5$ yr  (Murgia et al.\ 1999). 
It is believed that the smallest CSS and GPS objects, christened the 
compact symmetric objects, or CSOs, evolve to the medium
symmetric objects (MSOs) as the jets traverse outwards, and later on to the standard 
FRII radio sources (e.g.\ Carvalho 1985; Fanti et al. 1995;
Readhead et al. 1996a,b; De Young 1997; O'Dea 1998 and references therein;
Snellen et al. 2000; Perucho \& Mart\'{\i} 2002).  

\begin{figure*}
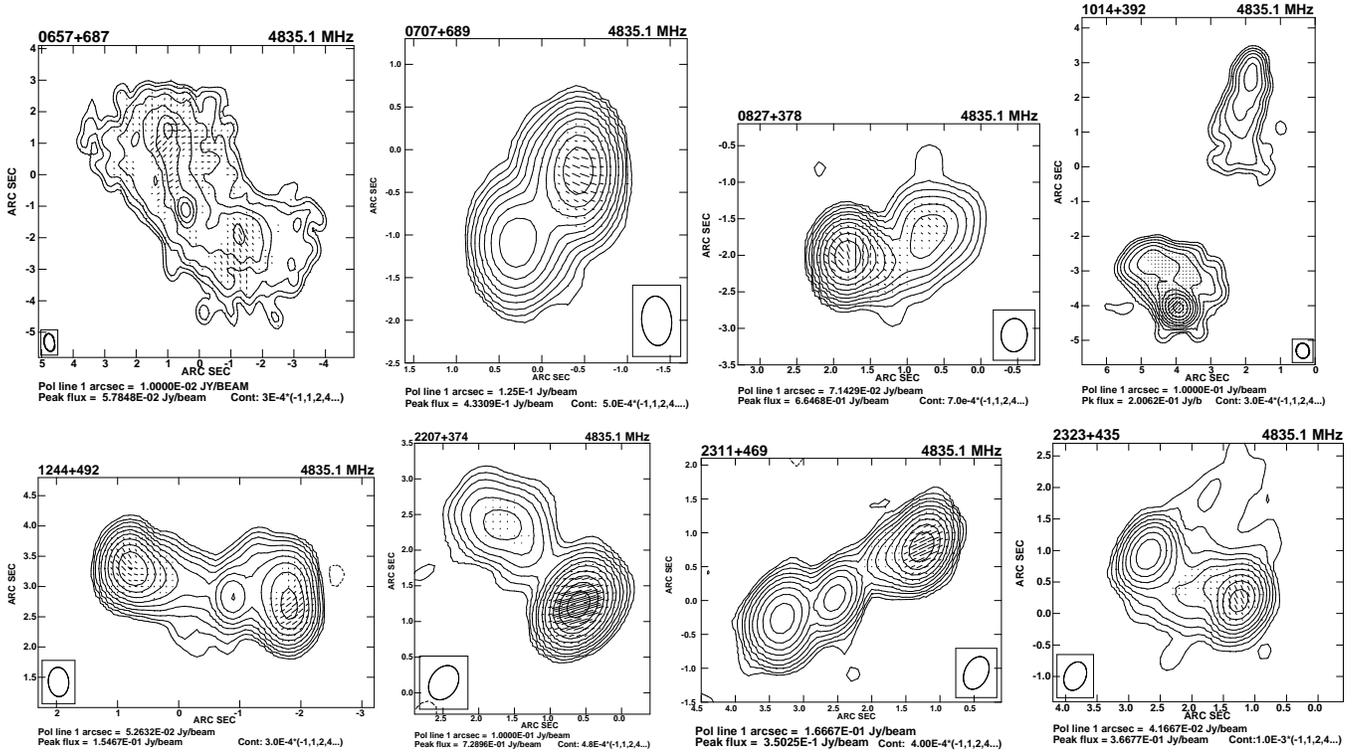

\vbox{
\hbox{
  \psfig{file=0657+68.PS.new,width=1.85in}
  \psfig{file=0707+68.PS.new,width=1.7in}
  \psfig{file=0827+37.PS.new,width=1.8in,angle=-90}
  \psfig{file=1014+39.PS.new,width=1.4in}
     }
\hbox{
  \psfig{file=1244+49.PS.new,width=1.95in,angle=-90}
  \psfig{file=2207+37.PS.new,width=1.4in}
  \psfig{file=2311+46.PS.new,width=1.80in,angle=-90}
  \psfig{file=2323+43.PS.new,width=1.75in}
    }
}
\caption[]{Images of the CSS sources at 4835 MHz with the polarization
E-vectors superimposed on the total-intensity contours. 
The peak brightness and contour levels in units of Jy/beam are given below each image.
}
\end{figure*}

An interesting related question is the fuelling of these young radio sources,
possibly due to the infall of gas to the central regions due to interacting 
companions and mergers. Detailed calculations  (e.g.\ Hernquist \& Mihos 1995)
showed that following a merger the infall of interstellar
matter into the central few hundred pcs takes place over a time scale of 
$\approx$10$^8$ yr. In such a situation, it should be possible to find evidence
of this infalling material by following the evolution of the radio components in these 
CSS sources as they advance outwards and interact with this material.
Their sizes and structures often appear
to be affected by the ambient gas in the central regions of the host galaxies  
(e.g. Spencer et al. 1991; Schilizzi et al. 2000),
suggesting strong dynamical interaction with the external medium.  The CSSs
tend to be more  asymmetric in both the brightness and location of the outer radio components
compared with the larger sources
(cf. Saikia et al. 1995, 2001; Arshakian \& Longair 2000). In a recent study of a sample of 
CSSs from the 3CR, S4, B2 
and B3 samples it has been shown that the CSSs exhibit large
brightness asymmetries with the flux density ratio for the opposing lobes being $>$5 for 
$\sim$25 per cent of the objects,
compared with only $\sim$5 per cent for the objects of larger size (Saikia et al. 2002).
The authors estimate the sizes of the clouds responsible for these asymmetries to
be $\sim$3 to 7 kpc, similar to those of dwarf galaxies (Swaters 1999), and speculate that such clouds
might be responsible for the infall of gas which fuels the radio source.

Evidence of this gas may also be probed via polarization measurements (e.g. Saikia et al.
1987; van Breugel et al. 1992; Mantovani et al. 1994, 2002; Akujor \& Garrington 1995; 
L\"{u}dke et al.  1998).  In a double-lobed 
radio source,  the component interacting with the dense gas should normally 
exhibit a higher rotation measure (RM)
as well as depolarize more rapidly than the component advancing through
the more tenuous medium. Although there have been relatively few studies of
RM determinations in CSS and GPS objects, partly due to their
low levels of polarization, such studies have sometimes revealed striking
differences in the RM on opposite sides of the nucleus or high RM in regions 
where the jets bend sharply. For example, 
polarization observations of 3C147  show huge differential RMs
between the two oppositely directed lobes, suggesting their evolution in an 
asymmetric environment (Junor et al.\ 1999). The quasar 3C216 has RMs in excess of
2000 rad m$^{-2}$ in the arc of emission $\sim$140 mas from the core where the jet
bends sharply (Venturi \& Taylor 1999). The CSS quasar 3C43 has also been
shown to have a high RM in the region where the jet bends sharply (Cotton et al. 2003). 

In this paper, we use the polarization asymmetry, r$_{\rm m}$, defined to be the ratio of the 
degree of scalar polarization of the lobes on opposite 
sides of the nucleus at cm wavelengths ($\lambda$2 to 6 cm) to probe ambient gas asymmetries 
for both CSS and larger 
FRII radio sources. We present polarization observations of double-lobed CSS objects,
and show that the CSS sources do tend to exhibit a higher polarization asymmetry than the
larger FRII radio sources. We also show that while the polarization asymmetry for CSS objects does
not depend on redshift, r$_{\rm m}$ increases with redshift for the more extended FRII sources. 
The implications of these results are discussed.
  
\section{The sample of sources}
The polarization of the lobes of the well-resolved, double-lobed CSS sources with
a measured redshift and a size $\leq$20 kpc from the S4 survey (Saikia et al. 2001)
were determined from Very Large Array A-array observations at 4835 MHz. 
The data were calibrated and reduced using standard imaging and
CLEANing procedures in the AIPS (Astronomical Image Processing
System) package. All flux densities are on the Baars et al. (1977) scale. Excluding 
well-studied sources, such as 3C216 and 3C295, the images for the remaining sources are
presented in Figure 1, and some of their parameters are 
listed in Table 1, which is arranged as follows.
Col. 1: source name in the IAU format using the B1950 coordinates; 
Cols. 2 to 4: the major and minor axes of the restoring 
beam in arcsec and its
position angle (PA) in degrees; Col. 5: the rms noise in the 
images in units of $\mu$Jy/beam; Col. 6: component designation where C denotes the core
or nuclear component; Col. 7: component flux densities, S, 
in units of mJy determined by integrating over a box around the component, except for the
core components where the peak values are listed;
Col. 8: the degree of polarization of the component, m.  We estimate m following the procedure
of Fanti et al. (2002), by blanking out points where the polarized intensity is $<3\sigma$ in 
order to avoid spurious contributions of low signal-to-noise ratio polarization data. 
A fractional polarization
image was then produced by excluding the total-intensity points which are also $<3\sigma$; m
represents the average value of fractional polarization while the upper limits represent about 3
times the error. The polarized intensity images were examined to ensure that a few
unblanked pixels at the outer edges of a source do not lead to spurious values of m.

%
%
 
\begin{table}
{\bf Table 1.} The observational parameters and derived properties \\

\begin{tabular}{l rrr c c rr }
\hline
Source & \multicolumn{3}{c}{Beam size} & rms & Cp & S & m \\
       &  $^{\prime\prime}$ & $^{\prime\prime}$ & $^\circ$ & $\mu$Jy/b   & & mJy   & \% \\
\hline
0657+68 & 0.56 & 0.33 &   9 &  50 &  S & 182 &         17.73 \\
        &      &      &     &     &  C &  56 &   $\lapp$0.50 \\
        &      &      &     &     &  N & 262 &         21.53 \\  
0707+68 & 0.58 & 0.35 &   5 & 165 &  W & 555 &          3.54 \\
        &      &      &     &     &  E & 151 &   $\lapp$0.55 \\
0827+37 & 0.46 & 0.35 & 177 & 110 &  W &  92 &          9.23 \\
        &      &      &     &     &  E & 766 &          5.46 \\
1014+39 & 0.44 & 0.37 &  10 &  75 &  N &  79 &   $\lapp$0.90\\
        &      &      &     &     &  S & 421 &         11.71 \\
1244+49 & 0.48 & 0.33 &   4 &  45 &  W & 229 &          5.98 \\
        &      &      &     &     &  C &  40 &   $\lapp$0.50 \\
        &      &      &     &     &  E & 278 &          5.53 \\
2207+37 & 0.51 & 0.38 & 148 & 110 &  W & 779 &          8.41 \\
        &      &      &     &     &  E &  53 &          5.82 \\
2311+46 & 0.51 & 0.34 & 155 & 150 &  W & 420 &         10.75 \\
        &      &      &     &     &  C &  75 &   $\lapp$0.75 \\ 
        &      &      &     &     &  E & 229 &   $\lapp$0.50 \\
2323+43 & 0.47 & 0.34 & 156 &  70 &  W & 544 &          1.94 \\
        &      &      &     &     &  E & 360 &   $\lapp$0.45 \\
\hline 
\end{tabular}
\end{table}


In addition to our observations, we have compiled from the literature the polarization
information of the lobes of all CSSs which satisfy our selection criteria. Those sources 
with a high limit to the degree of polarization of one of the lobes, or with limits for
both the lobes have not been included since it is difficult to get meaningful estimates
of r$_{\rm m}$ for these objects. The final sample of 30 objects is
listed in Table 2 which is arranged as follows.
Col. 1: source name in the IAU format using the B1950 co-ordinates; Col. 2: an alternative name;
Col. 3: optical classification where G denotes a galaxy and Q a quasar;
Col. 4: redshift; Column 5: the projected linear size in kpc 
measured from the outermost peaks of radio emission; 
Col. 5: the wavelength of observation in cm; 
Col. 6: the ratio of the degree of polarization of the oppositely-directed lobes, r$_{\rm m}$;
Col. 7: references used for estimating these values. 

%
%

\begin{table}
{\bf Table 2.} The sample of CSS objects \\

\begin{tabular}{l l c c c c r c}
\hline
Source    & Alt.    &  O   &    z   &  $l$  & $\lambda$ &   r$_{\rm m}$ & Rf.  \\
          & name    &      &        &  kpc  &    cm     &               &      \\
\hline 
0127+233  & 3C43    &  Q   &  1.459 &  16   &   3.6     &         1.8     &  1    \\
0221+276  & 3C67    &  G   &  0.310 &  7.6  &   3.6     & $\gapp$27.8     &  1    \\
0518+165  & 3C138   &  Q   &  0.759 &  3.0  &   6.0     & $\gapp$40.0     &  3    \\
0538+498  & 3C147   &  Q   &  0.545 &  2.5  &   2.0     &        15.6     &  2    \\
0657+687  & 4C68.07 &  G   &  0.110 &  5.6  &   6.0     &         1.2     &  P    \\
0707+689  & 4C68.08 &  Q   &  1.141 &  6.3  &   6.0     &  $\gapp$6.4     &  P    \\
0827+378  & 4C37.24 &  Q   &  0.914 &  5.9  &   6.0     &         1.7     &  P    \\
0858+292  & 3C213.1 &  G   &  0.194 &  13   &   3.6     &         1.1     &  1    \\     
0906+430  & 3C216   &  Q   &  0.668 &  7.5  &   3.6     &         5.5     &  1    \\
1005+077  & 3C237   &  G   &  0.877 &  6.9  &   3.6     &  $\gapp$3.3     &  1    \\
1014+392  & 4C39.29 &  G   &  0.206 &  16   &   6.0     & $\gapp$13.0     &  P    \\
1153+317  & 4C31.38 &  Q   &  1.577 &  5.6  &   6.0     &         1.5     &  4    \\ 
1203+645  & 3C268.3 &  G   &  0.371 &  4.6  &   3.6     &        21.4     &  1    \\
1244+492  & 4C49.25 &  G   &  0.206 &  6.0  &   6.0     &         1.1     &  P    \\
1250+568  & 3C277.1 &  Q   &  0.321 &  5.2  &   3.6     &        11.5     &  1    \\
1350+316  & 3C293   &  G   &  0.045 &  2.2  &   3.6     &        25.6     &  1    \\
1409+524  & 3C295   &  G   &  0.461 &  18   &   3.6     &         2.4     &  1    \\
1416+067  & 3C298   &  Q   &  1.439 &  9.0  &   3.6     & $\gapp$21.2     &  1    \\
1443+773  & 3C303.1 &  G   &  0.267 &  4.7  &   6.0     &         3.0     &  3    \\
1447+771  & 3C305.1 &  G   &  1.132 &  13   &   3.6     &         1.9     &  1    \\
1458+718  & 3C309.1 &  Q   &  0.904 &  11   &   3.6     &         2.1     &  1    \\
1517+204  & 3C318   &  G   &  0.752 &  4.9  &   3.6     &  $\gapp$7.7     &  1    \\
2207+374  & 4C37.65 &  Q   &  1.493 &  9.6  &   6.0     &         1.4     &  P    \\
2247+140  & 4C14.82 &  Q   &  0.237 &  0.5  &   2.0     &        10.8     &  4    \\
2248+712  & 3C454.1 &  G   &  1.841 &  10   &   3.6     &         1.2     &  1    \\
2249+185  & 3C454   &  Q   &  1.757 &  4.2  &   6.0     &        23.1     &  3    \\ 
2252+129  & 3C455   &  G   &  0.543 &  14   &   3.6     &         6.8     &  1    \\
2311+469  & 4C46.47 &  Q   &  0.742 &  11   &   6.0     & $\gapp$21.5     &  P    \\           
2314+038  & 3C459   &  G   &  0.220 &  19.5 &   6.0     &        19.8     &  5    \\
2323+435  & OZ+438  &  G   &  0.145 &  2.8  &   6.0     &  $\gapp$4.3     &  P    \\
\hline
\end{tabular}
References:
1: Akujor \& Garrington 1995; 2: Junor et al. 1999; 
3: L\"{u}dke et al. 1998; 4: van Breugel et al. 1984; 
5: Thomasson et al. 2003; P: Present work
\end{table}


For comparison with a sample of sources of larger linear sizes but observed at a similar wavelength,
we have considered the $\lambda$6 cm observations of the sample of 47 sources by 
Garrington et al. (1991) to investigate the Laing-Garrington effect (Laing 1988;
Garrington et al. 1988). This sample has a median
linear size of 100 kpc and a median redshift of 1, with 40 of the sources being associated with
quasars and 7 with radio galaxies.  It is relevant to note that this sample was
selected due to the presence of a radio jet and its degree of polarization asymmetry is likely to
be higher than for a randomly selected sample, or one which has a larger proportion of galaxies
(Holmes 1991). This selection effect will tend to diminish any possible 
difference in the degree of polarization asymmetry between the CSS and larger sources. 

\begin{figure}
\vbox{
  \psfig{file=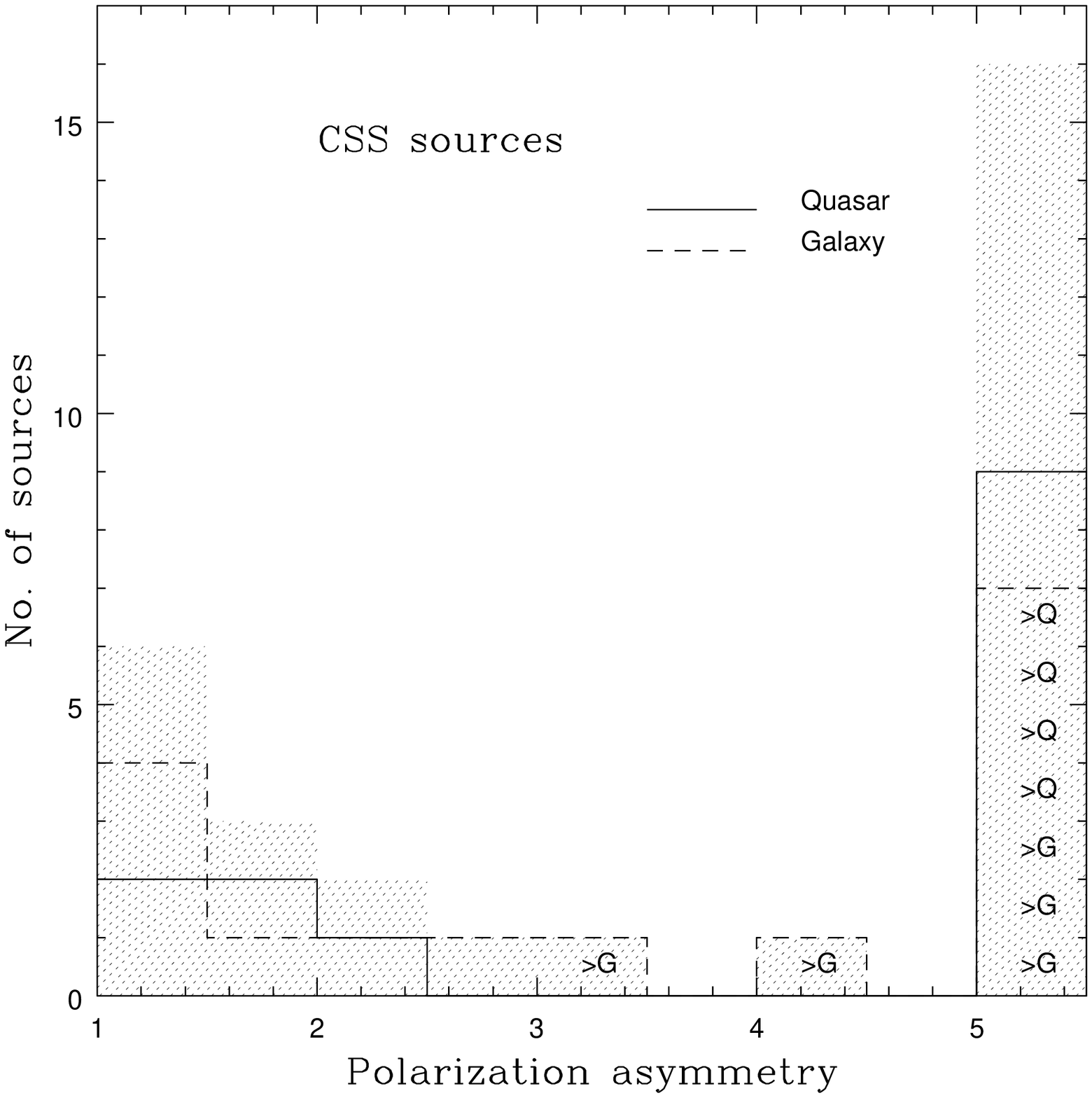,width=2.7in}
  \psfig{file=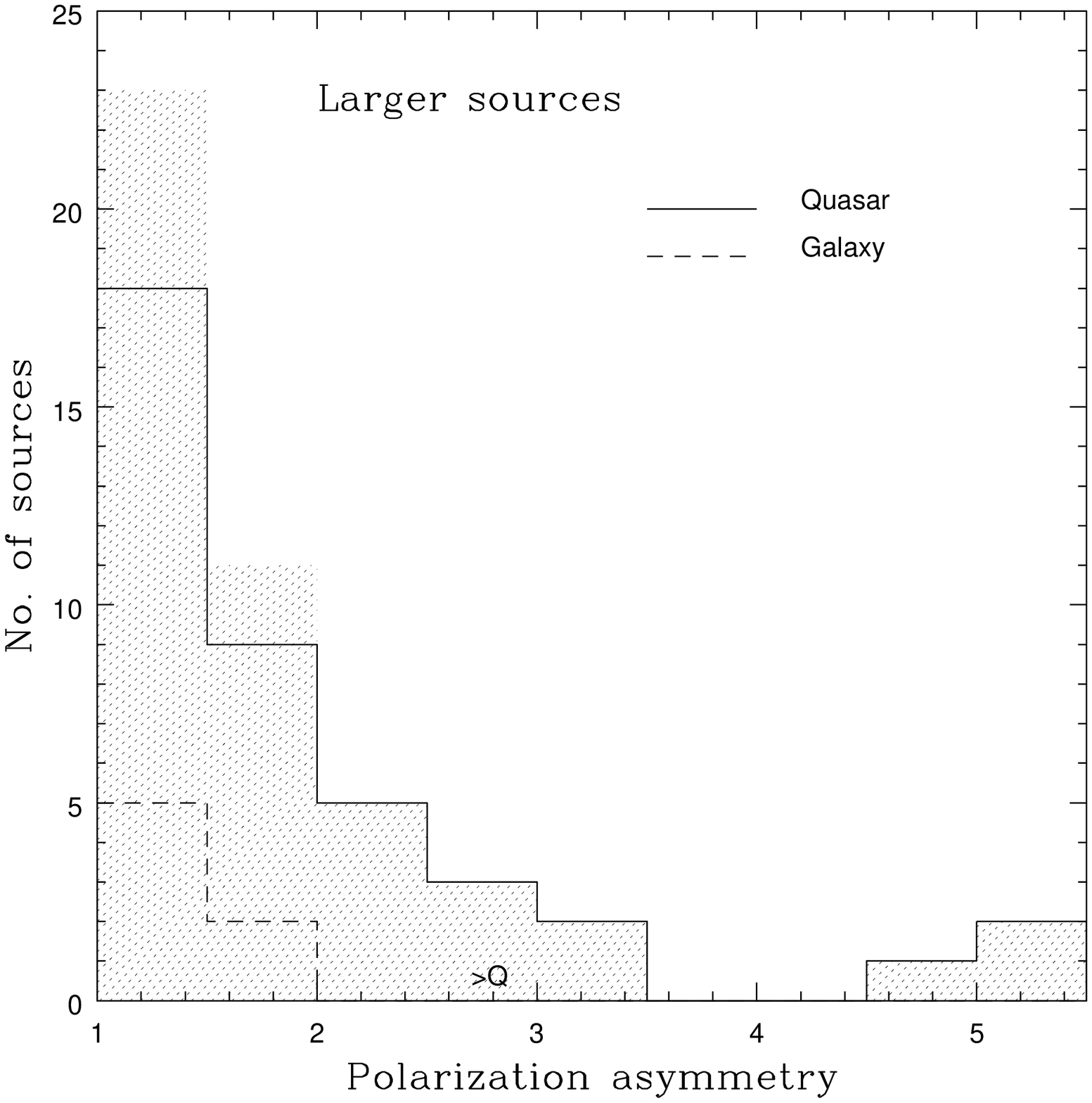,width=2.7in}
   }
\caption[]{The distributions of the polarization asymmetry parameter,
r$_{\rm m}$ for the CSS and larger sources.
All the sources with r$_{\rm m}$ $>$5 have been placed in the last bin. 
The distributions for the entire sample are shown shaded.
}
\end{figure}

\section{Results and discussion}
The distribution of r$_{\rm m}$ for the entire sample of CSS objects (Fig. 2),
show that $\sim$50 \% of the sources 
appear to have r$_{\rm m}$$>$5, yielding a median value of $\sim$5.
There is a trend for the quasars to be more asymmetric than the 
galaxies which could be due to effects of orientation such as unresolved jets in some of
the lobes.  Also, there is a suggestion that the smallest CSS objects, with a linear size,
say $<$5 kpc, tend to exhibit a higher degree of polarization asymmetry, compared with the
larger CSS objects. 

The distribution of r$_{\rm m}$ for the control sample of larger sources is also shown in Fig. 2.
The median value for the entire sample is $\sim$1.5, with no significant difference between those
with sizes either larger or smaller than the median value of 100 kpc. The median value of r$_{\rm m}$
for the 7 galaxies in the sample is also similar.
Only 2 of the objects ($\sim$5 \%) 
exhibit a value of r$_{\rm m}$$>$5. These two, namely 0802+10 (3C191) and 1857+566, 
as well as 0017+15 (3C9) with r$_{\rm m}$$\sim$4.7 are dominated by prominent one-sided 
jets with a high degree of polarization (L\"{u}dke et al. 1998; 
Saikia et al. 1983; Bridle et al. 1994). Such jets are highly dissipative with a more prominent
hot-spot on the opposite side
and are possibly intrinsically asymmetric in addition to effects of relativistic beaming (Saikia 1984).

The difference in the r$_{\rm m}$ distributions for the CSS and larger sources is very striking
and highly significant. A 
Kolmogorov-Smirnov test shows the two populations to be different at a significance
level of $>$99.9\%. We have also confirmed this result by considering CSS and larger sources
observed with a similar number of resolution elements along the source axes (median value $\sim$10),
by excluding 4 CSS and 5 larger objects which have been observed with less than 4 resolution elements.
The tendency for CSS objects to have large values of r$_{\rm m}$ is also not due to one side
of these objects being dominated by a prominent one-sided jet. Of the 16 sources with r$_{\rm m}>$5,
approximately three-fourths of these have either no detected jets or only weak jets which 
do not contribute significantly to the 
polarized flux density of the lobes. Further, 7 of these 16 objects with r$_{\rm m}>$5 are associated 
with galaxies, which are expected to be at large angles to the line of sight and have at best weak
radio cores and jets. This is consistent with the observations of these sources. It is also 
interesting to note that while some of the jets
in CSS objects are strongly polarised at cm wavelengths such as in 3C138, there are also cases
where the jets exhibit no significant polarised flux density as in 3C186, possibly due to 
Faraday depolarisation effects (L\"{u}dke et al. 1998). 
A large sample of CSS objects which has been observed in both total intensity
and linear polarization is the B3-VLA sample with S(408) $\geq$ 0.8 Jy (Fanti et al. 2001). 
We have done a similar analysis for their sample using their listed average values
of fractional polarization for the strongest components on opposite sides for the double-lobed sources.
We again find a similar trend;  the median value of the fractional polarization ratio is $\sim$3 with
approximately a third of the objects having a value $>$5. 

This higher degree of polarization asymmetry in CSS objects could be due to 
interactions with dense clouds of gas which possibly fuel the radio source. 
The interactions of the jets with such clouds
could either compress and shear magnetic fields, increasing the degree of polarization, or 
depolarise the emission due to Faraday effects. Both effects could contribute towards increasing
the degree of polarization asymmetry of the oppositely-directed lobes. One would therefore not
expect a significant
correlation between the polarization and flux density asymmetries of the oppositely-directed 
lobes. The data also does not show a significant relation between these two parameters.

These dense clouds
with sizes possibly similar to those of dwarf galaxies, are embedded in the 
large-scale halos with typical core radii of $\sim$100 kpc which cause the 
Laing-Garrington effect. It is worth noting that these halos 
will cause only marginal polarization asymmetry on the scale of the CSS objects (Garrington \&
Conway 1991). Also, sometimes it is the jet side which is more strongly depolarized
and exhibit low levels of polarization as seen, for example, in 3C459 (Thomasson et al. 2003). 

It is also of interest to enquire whether the polarization asymmetries 
depend on cosmic epoch because of the larger incidence of interactions and
mergers in the past, as seen in Hubble Space Telescope studies of distant galaxies (cf.
Abraham et al. 1996, Brinchmann et al. 1998, Ellis et al. 2000). The r$_{\rm m}$$-$redshift
diagram for the CSS objects (Fig. 3)
shows no significant dependence of r$_{\rm m}$ on redshift, although polarization observations
of a larger number of sources at higher redshifts would be useful to confirm that this is indeed the
case.  The absence of a significant relationship suggests that although interactions and  
mergers may increase with redshift 
globally, the density asymmetries in the central regions of these active galaxies on the 
scale of the CSS objects, which are young and possibly still being
fuelled by the infall of gas, are similar at different redshifts. 

The r$_{\rm m}$$-$redshift diagram for the larger sources (Fig. 3) shows a clear trend for
the polarization asymmetry to be higher at larger redshifts. Dividing the entire sample into
the low- and the high-redshift ones at the median redshift of 1, the median value of r$_{\rm m}$
is $\sim$1.4 for the low-redshift sources and $\sim$2 for the high-redshift objects. Leaving out
the three objects with r$_{\rm m}$ $>$4.5, reduces the median value for the high-redshift
objects to $\sim$1.8, which is still significantly higher than for the low-redshift sample.
A Kolmogorov-Smirnov test shows the low- and high-redshift distributions to be different at 
a significance level of $>$99 per cent. Excluding the three outlying jet-dominated sources
with r$_{\rm m}$ $>$4.5 reduces the significance level to $\sim$98 per cent. The number of
resolution elements along the source axes for these sources are similar at different redshifts. 
The tendency for the high-redshift objects to exhibit
a higher degree of polarization asymmetry could be due to interactions with companions on these
scales. This would be consistent with the HST results mentioned earlier, and is also reminiscent
of higher redshift radio sources appearing more distorted possibly due to interactions with 
companions (Barthel \& Miley 1988).

\section{Conclusions}
We present linear polarization observations of a sample
of CSS sources, and show that the 
CSS sources are more asymmetric in the polarization of the outer lobes compared
with those of the more extended ones.  This could be possibly due
to interaction of the jets with infalling material, which fuels the radio source.
The polarization asymmetry of the lobes of CSS sources show no significant dependence on 
redshift, suggesting that the environments on the
CSS scales are similar at different redshifts. However, the polarization asymmetry of the
oppositely-directed lobes increases with redshift for the more extended sources,
possibly reflecting the higher incidence of mergers and interactions in the past.

\begin{figure}
\vbox{
  \psfig{file=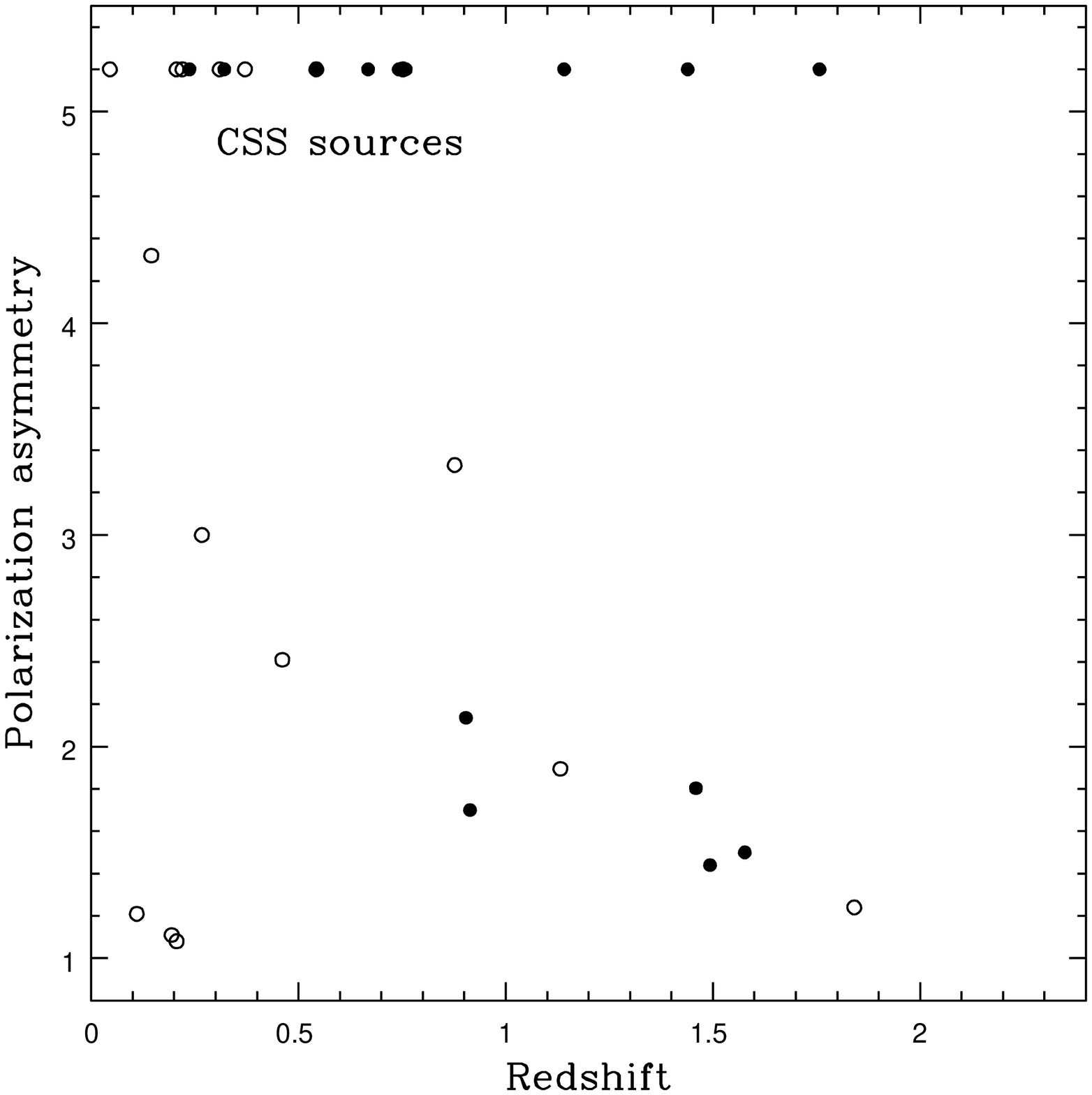,width=2.7in}
  \psfig{file=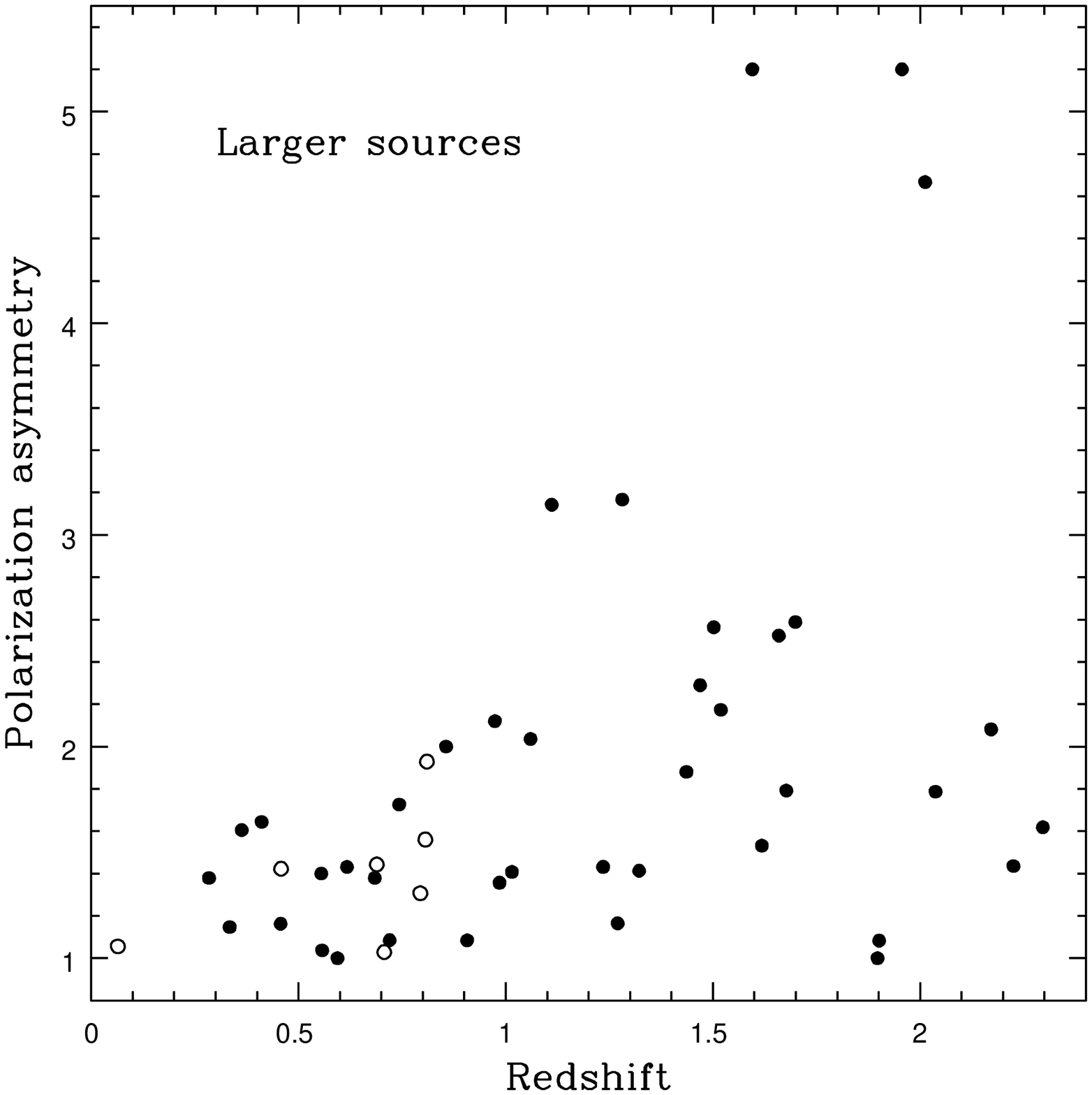,width=2.7in}
   }
\caption[]{The r$_{\rm m}$$-$redshift diagram for the CSS
and larger sources. Those with r$_{\rm m}$$>$5 are shown at r$_{\rm m}$$=$5.2. The open
and filled circles represent galaxies and quasars respectively. 
}
\end{figure}

\begin{acknowledgements}
We thank Carla Fanti, the referee for several valuable comments and suggestions, and also for
her very prompt refereeing of the paper. We also thank our colleagues and collaborators for 
their comments.
The National Radio Astronomy Observatory is a facility of the National Science Foundation
operated under co-operative agreement by Associated Universities, Inc.  We have made use of the
NASA/IPAC Extragalactic Database (NED), which is operated by the Jet Propulsion Laboratory,
California Institute of Technology under contract with the National Aeronautics and Space
Administration. 
\end{acknowledgements}


\begin{thebibliography}{}

\bibitem[]{} Abraham, R.G., Tanvir, N.R., Santiago, B.X., et al. 1996, MNRAS, 279, L47
\bibitem[]{} Akujor, C.E., \& Garrington, S.T. 1995, A\&AS, 112, 235
\bibitem[]{} Arshakian, T.G., \& Longair, M.S. 2000, MNRAS, 311, 846
\bibitem[]{} Baars, J.W.M., Genzel, R., Pauliny-Toth, I.I.K., \& Witzel, A. 1977, 
             A\&A, 61, 99
\bibitem[]{} Barthel, P.D., \& Miley, G.K. 1988, Nature, 333, 319
\bibitem[]{} Bridle, A.H., Hough, D.H., Lonsdale, C.J., Burns, J.O., \& Laing, R.A. 1994, AJ, 108, 766  
\bibitem[]{} Brinchmann, J., Abraham, R., Schade, D., et al. 1998, ApJ, 499, 112
\bibitem[]{} Carvalho, J.C. 1985, MNRAS, 215, 463
\bibitem[]{} Cotton, W.D., Spencer, R.E., Saikia, D.J., \& Garrington, S.T. 2003, A\&A, submitted
\bibitem[]{} De Young, D.S. 1997, ApJ, 490, 55L
\bibitem[]{} Ellis, R.S., Abraham, R.G., Brinchmann, J., \& Menanteau, F. 2000, 
             A\&G, No. 2, 10
\bibitem[]{} Fanti, C., Fanti, R., Dallacasa, D., et al. 1995, A\&A, 302, 317
\bibitem[]{} Fanti, C., Pozzi, F., Dallacasa, D., et al. 2001, A\&A, 369, 380
\bibitem[]{} Garrington, S.T., Leahy, J.P., Conway, R.G., \& Laing, R.A.  
             1988, Nature, 331, 147
\bibitem[]{} Garrington, S.T., Conway, R.G., \& Leahy, J.P. 1991, MNRAS, 250, 171
\bibitem[]{} Garrington, S.\ T., \& Conway, R.\ G.  1991, MNRAS, 250, 198
\bibitem[]{} Hernquist, L., \& Mihos, J.\ C. 1995, ApJ, 448, 41
\bibitem[]{} Holmes, G.F. 1991, PhD thesis, University of Manchester 
\bibitem[]{} Junor, W., Salter, C.J., Saikia, D.J., Mantovani, F.M., \& Peck, A.B. 1999, 
             MNRAS, 308, 955
\bibitem[]{} Laing, R.\ A., 1988, Nature, 331, 149
\bibitem[]{} L\"{u}dke, E., Garrington, S.T., Spencer, R.E., et al. 1998, MNRAS, 299, 467
\bibitem[]{} Mantovani, F., Junor, W., Fanti, R., Padrielli, L., \& Saikia, D.J. 1994, 
             A\&A, 292, 59
\bibitem[]{} Mantovani, F., Junor, W., Ricci, R., et al. 2002, A\&A, 389, 58
\bibitem[]{} Murgia, M., Fanti, C., Fanti, R., et al.  1999, A\&A, 345, 769
\bibitem[]{} O'Dea, C.P. 1998, PASP, 110, 493
\bibitem[]{} Perucho, M., \& Mart\'{\i}, J.\ M.  2002, ApJ, 568, 639
\bibitem[]{} Polatidis, A.G., \& Conway, J.E. 2003, PASA, 20, in press.
\bibitem[]{} Readhead, A.C.S., Taylor, G.B., Xu, W., et al. 1996a, ApJ, 460, 612
\bibitem[]{} Readhead, A.C.S., Taylor, G.B., Pearson, T.J., \& Wilkinson, P.N. 1996b, 
             ApJ, 460, 634
\bibitem[]{} Saikia, D.J. 1984, MNRAS, 208, 231
\bibitem[]{} Saikia, D.J., Shastri, P., Cornwell, T.J., \& Banhatti, D.G. 1983, MNRAS, 203, 53P
\bibitem[]{} Saikia, D.J., Singal, A.K., \& Cornwell, T.J. 1987, MNRAS, 224, 379 
\bibitem[]{} Saikia, D.J., Jeyakumar, S., Wiita, P.J., Sanghera, H.S., \& Spencer, R.E. 1995,
             MNRAS, 276, 1215 
\bibitem[]{} Saikia, D.J., Jeyakumar, S., Salter, C.J., et al. 2001, MNRAS, 321, 37
\bibitem[]{} Saikia, D.J., Thomasson, P., Spencer, R.E., et al. 2002, A\&A, 391, 149
\bibitem[]{} Schilizzi, R.T., Tschager, W., Snellen, I.A G., et al. 2000, AdSpR, 26, 709 
\bibitem[]{} Snellen, I.A.G., Schilizzi, R.T., Miley, G.K., et al. 2000, MNRAS, 319, 445
\bibitem[]{} Spencer, R.E., Schilizzi, R.T., Fanti, C., et al. 1991, MNRAS, 250, 225 
\bibitem[]{} Swaters, R.A. 1999, Ph.D. thesis, Rijksuniversiteit Groningen
\bibitem[]{} Taylor, G.B., Marr, J.\ M., Pearson, T.J., \& Readhead, A.C.S.
           2000, ApJ, 541, 112
\bibitem[]{} Thomasson, P., Saikia, D.J.,  \& Muxlow, T.W.B., 2003, MNRAS, in press 
\bibitem[]{} van Breugel, W.J.M., Miley, G., \& Heckman, T. 1984, AJ, 89, 5
\bibitem[]{} van Breugel, W.J.M., Fanti, C., Fanti, R., et al. 1992, A\&A, 256, 56
\bibitem[]{} Venturi, T., \& Taylor, G.B. 1999, AJ, 118, 1931


\end{thebibliography}
\end{document}